\documentclass{emulateapj}

\usepackage{apjfonts}

 \font\sevenrm=cmr7 scaled 1000
\font\sixrm=cmr6 scaled 1000  

\slugcomment{Accepted to ApJL}

\begin{document}
\title{Differences in the AGN Populations of Groups and Clusters: Clues to AGN Evolution}

\shorttitle{}

\shortauthors{Shen et al. }

\author{Yue Shen\altaffilmark{1,2}, John S. Mulchaey\altaffilmark{2},
Somak Raychaudhury\altaffilmark{3}, Jesper
Rasmussen\altaffilmark{3} and Trevor J. Ponman\altaffilmark{3}}

\altaffiltext{1}{Department of Astrophysical Sciences, Princeton
University, Princeton, NJ 08544; yshen@astro.princeton.edu}
\altaffiltext{2}{Carnegie Observatories, 813 Santa Barbara Street,
Pasadena, CA 91101; mulchaey@ociw.edu} \altaffiltext{3}{School of
Physics and Astronomy, University of Birmingham, Edgbaston,
Birmingham B15 2TT, UK}

\begin{abstract}
We combine optical and X-ray data for eight low redshift ($z\sim
0.06$) poor groups of galaxies from the {\it XI }({\it XMM/IMACS})
Groups Project to study the AGN population in the group
environment. Among $\sim 140$ group members, we identify five AGN
based on their optical emission lines. None of these
optically-selected AGN are detected by {\it XMM-Newton}. One
additional AGN is discovered in the {\it XMM-Newton} observations.
This X-ray detected AGN,
which has no obvious AGN emission line signatures in its optical
spectrum, is a member of the only X-ray luminous group in our
sample. The lack of a significant population of X-ray bright, but
optically dull AGN among less dynamically evolved groups is in
stark contrast to the large fraction of such objects in rich
clusters of galaxies (Martini et al. 2006).
%We suggest that this
%result can be explained if the AGN accretion rates in groups are
%higher than in rich clusters.
%Based on the differences in the AGN
%populations of groups and clusters,
We suggest this result can be explained by a physical scenario for AGN
accretion evolution: AGN activity is initially triggered by galaxy
merging, leading to a high accretion rate and an
optically dominant phase (via thin disk accretion). %Most of its
%supermassive black hole (SMBH) mass is assembled during this time.
As the accretion rate drops in time, the AGN gradually enters an
%optical-dull but X-ray bright
X-ray dominant low-accretion phase (via a radiative inefficient
accretion flow).
%The change of accretion rate and disk accretion mechanism
%occurs on a typical timescale of $\sim$ a few $10^7$ yr,
%approximately the lifetime for the early optically-bright phase.
In this picture, optical- and X-ray-selected AGN are the same
population of supermassive black holes observed at different
epochs. Within the context of this scenario, the majority of AGN
in poor groups are in the high-accretion optically dominant phase,
while the AGN population in rich clusters is mostly in the
low-accretion X-ray dominant phase.
%rather than two different populations.
%Observational implications
%of this model are discussed.
\end{abstract}

\section{Introduction}
Active Galactic Nuclei (AGN) play a key role in modern cosmology
and hierarchical galaxy formation. Powered by gas accretion onto a
central supermassive black hole (SMBH; e.g., Salpeter 1964;
Lynden-Bell 1969; Rees 1984), AGN dynamically impact the formation
and evolution of host galaxies in a self-regulated way (e.g.,
Begelmen 2004 and references therein).
%Despite many advances made towards
%understanding AGN in the past few decades, many
%questions remain unanswered. %For example, constraints on the
%optical AGN lifetime are still uncertain by two orders of
%magnitude (see Martini 2004 for a review); the physics of AGN
%accretion processes is not very well understood; and the
%connection between SMBH growth and bulge formation is still not
%very clear.

Recently, X-ray surveys with {\it Chandra} and {\it XMM-Newton}
have revealed a population of AGN that have very different
properties from optically-selected AGN. In particular, X-ray AGN
have a higher spatial number density that peaks at a
lower redshift than that of optically-selected AGN (Barger et al.
2002; Fabian 2004; Hasinger, Miyaji \& Schmidt 2005). Also, X-ray
AGN cluster more strongly than optically-selected AGN (e.g.,
Basilakos et al. 2004; Yang et al. 2006; Puccetti et al. 2006).
Detailed studies of the connection between X-ray and optical AGN
are often hampered by the difficulty of obtaining sufficient
optical spectroscopy and X-ray data. Nevertheless, studies of
clusters of galaxies reveal the existence of a large number of
X-ray AGN that show little or no optical AGN signatures (Martini
et al. 2002, 2006).

Recent results on the small-scale excess of quasar-quasar and
quasar-galaxy pairs (Hennawi et al. 2006; Serber et al. 2006)
support models where AGN activity is triggered by gas fueling
during galaxy mergers (e.g. Kauffmann \& Haehnelt 2000). Given
their low velocity dispersions and high galaxy densities, groups
of galaxies provide an ideal environment for such mergers. As
groups constitute the most common galaxy associations, a
significant fraction of all AGN activity may be triggered in these
systems. In this Letter, we use optical and X-ray observations to
search for AGN in a sample of eight low-redshift groups taken from
the XI (XMM/IMACS) Groups Project (Rasmussen et al. 2006). These
redshift-selected groups were chosen to be representative of poor
groups at low redshift\footnote{The XI main group sample contains
25 groups taken from the Merchan \& Zandivarez (2002) catalog,
which satisfy: 1) redshift $z=0.060-0.063$; 2) velocity dispersion
$\lesssim 500{\rm\ km\ s^{-1}}$; 3) number of spectroscopically
confirmed 2dF member galaxies greater than 5. More details about
the sample selection can be found in Rasmussen et al. (2006).}.
All but one of these groups are weak or undetected in the X-ray
band, suggesting these systems are dynamically very young.
Therefore, the galaxies in these systems are likely encountering
the group environment for the first time. These groups are natural
places to study AGN which may have been recently triggered. We
describe the data and identification of AGN in \S\ref{sec:data}. A
comparison of the AGN population in our group sample with that in
rich clusters is provided in \S\ref{sec:comparison} and a physical
interpretation is presented in \S\ref{sec:unified_model}. We adopt
a flat $\Lambda$CDM cosmology with $\Omega_{M}=0.27$,
$\Omega_\Lambda=0.73$, and $H_0=72\ {\rm km\ s}^{-1}\ {\rm
Mpc}^{-1}$ throughout this Letter.

\section{Data Analysis and Results}\label{sec:data}

\begin{deluxetable}{lccccc}
\tablecolumns{6} \tablewidth{0pc} \tablecaption{Summary of group
properties\label{table:group}} \tablehead{Group & $\bar{z}$ &
N$_{\rm gal}$ & $\sigma_v$ & {\it XMM} ET & $L_{{\rm X,
limit}}^{0.3-8}$ \\ & & &(${\rm km\ s^{-1}}$) & (k s)& (${\rm erg\
s^{-1}}$)} \startdata
MZ 3849  &0.06054& 11 & 324$^{+75}_{-38}$ & 56.5& 8 $\times$ 10$^{39}$  \\
MZ 4577  &0.06227 &15 &229$^{+72}_{-35}$ & 4.4& 1 $\times$ 10$^{41}$\\
MZ 4592  &0.06142 &24 &212$^{+41}_{-26}$ &22.2 & 2.5 $\times$ 10$^{40}$\\
MZ 4940  &0.06200 & 8&66$^{+42}_{-12}$  &51.7 & 9 $\times$ 10$^{39}$\\
MZ 5293  &0.06210 & 7 &105$^{+36}_{-15}$ &19.8 & 2 $\times$ 10$^{40}$ \\
MZ 5383  &0.06030 & 22 &489$^{+75}_{-51}$ & 32.2& 1.5 $\times$ 10$^{40}$\\
MZ 9014  &0.06080 &23 &251$^{+60}_{-30}$ & 76.5&  6 $\times$ 10$^{39}$\\
MZ 10451 &0.06099 &29 &388$^{+51}_{-40}$ &36.8 & 1 $\times$
10$^{40}$ \smallskip
\enddata
\tablecomments{Optical group properties have been calculated using
all of the known members (including some below the magnitude cut
used for the AGN analysis). XMM exposure times refer to the total
summed over the three EPIC detectors. The X-ray limits correspond
to the 5$\sigma$ limit for a point source 13$'$ off-axis in energy
band $0.3-8$ keV.}
\end{deluxetable}

\subsection{IMACS data}
Multi-object spectroscopy of each group field was obtained using
the IMACS spectrograph on the Baade/Magellan telescope. The f/2
camera mode was used with the 300 lines mm$^{-1}$ grism, giving a
wavelength range of $3900-10000$ \AA\ and a dispersion of 1.34
\AA\ pixel$^{-1}$. Details of the object selection, exposure times
and data reduction can be found in Rasmussen et al. (2006). Galaxy
redshifts were obtained by cross-correlating the spectra with SDSS
galaxy templates and manually inspecting each fit to verify the
redshift. For a few bright objects that have not been targeted yet
with IMACS, we use spectra from the 2df Galaxy Redshift Survey
(Colless et al. 2001). Combining our IMACS spectra with the 2df
data, we are 100\% complete down to $M_{\rm R}=-20$ for all of the galaxies
in all eight groups
within a radius of 15$'$ ($\sim$ 1 Mpc) from the group
center (corresponding to the approximate XMM-Newton field of
view); 91\% complete down to $M_{\rm R}=-19$; and 74\% complete
down to $M_{\rm R}=-18$. We restrict our current analysis to group
members brighter than $M_{\rm R}=-18$. Group members are
identified using the technique described in Mulchaey et al.
(2006). The group properties are listed in Table 1.

A large fraction ($\gtrsim$ 60\%) of the group members show strong
emission lines. For each emission line object, a line fitting
procedure has been used to measure the line equivalent width (EW).
A Gaussian profile is assumed for the line and fit simultaneously
with the local continuum. The EW is calculated using the fitted
line area divided by the median value of the continuum within $\pm
3\sigma$ of the line center.

%Many ($>60\%$) group members show strong emission
%lines, with H$\alpha$ EW $>5$\AA\ , indicating intense ongoing
%star formation (e.g., Balogh et al. 2004). R$-$band images of the
%groups reveal clear evidence for galaxy encounters and
%interactions in many cases. Morphologically, most of the emission
%line galaxies are of late-type.

Traditionally, AGN have been distinguished from star forming
galaxies by considering the line flux ratio ${\rm [O\ {\sixrm
III}] \lambda 5007/H\beta}$ versus ${\rm [N\ {\sixrm II}] \lambda
6584/H\alpha}$ in the BPT diagram (Baldwin, Phillips \& Terlevich
1981; Kauffmann et al. 2003; Hao et al. 2005). Instead, we use
line EW ratios since our spectra are not flux calibrated. For
lines close in wavelength, the EW ratio is similar to the flux
ratio because the underlying continuum does not differ
significantly. %The original BPT diagram using line flux ratios and
%the alternative version using EW ratios are shown in Fig.
%\ref{BPT} for a sample of 20,000 galaxies from the SDSS, which
%demonstrates that EW ratios suffice for identifying AGN.
%We have compared the BPT diagrams using line flux ratios and using
%EW ratios for a sample of 20,000 galaxies drawn from the SDSS, and
%verified that EW ratios suffice for identifying AGN.

\begin{figure}
\centering
\includegraphics[width=0.5\textwidth]{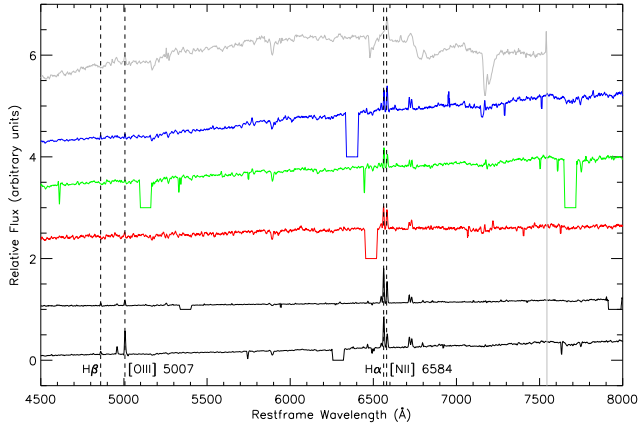}
\caption{Spectra of the five optical AGN (flux properly scaled):
two in MZ 4577 ({\it black}), one in MZ 4592
    ({\it red}), one in MZ 4940 ({\it green}), one in MZ 5293 ({\it blue}). Also shown is
    the X-ray detected AGN in MZ 10451 ({\it gray}).
    }
\label{AGN_spec}
\end{figure}

We restrict our analysis to emission lines with EWs $>$ 2\ \AA.
When all four emission lines can be measured in the spectrum, we
use eqn. 5 in Kewley et al. (2001)
%, which is a more strict criterion
%than the one used in Kauffmann et al. (2003),
to separate AGN from
star-forming galaxies. For some objects, the lines of H$\beta$
and/or [OIII] cannot be measured because they are either too weak
or they fall on bad parts of the spectrum due to bad pixels or
gaps in the IMACS CCD mosaic. For these objects, we use ${\rm EW\
[N\ {\sevenrm II}]}\lambda 6584/{\rm EW\ H\alpha}>0.63$ to
identify AGN (e.g., Kauffmann et al. 2003). Adopting these
definitions, we identify five AGN (out of a total number of $\sim
140$ galaxies) in the group sample. For the remainder of this
Letter, we refer to these AGN as optical
AGN. %\footnote{But note that our AGN selection is based on emission
%line features, not on blue/UV excess.}
The spectra of the five AGN are shown in Fig. \ref{AGN_spec}. The
objects with weak H$\beta$ and [OIII] occur in inclined host
galaxies and thus extinction in the host is probably responsible
for the weakness of these features. Optical images of the AGN
indicate that they reside in late-type galaxies.
%For some of the group members, the
%spectra are too noisy to measure line EWs. Thus, some optical AGN
%may be missing from our sample particularly among the lowest
%luminosity galaxies.

\subsection{XMM-Newton Data}
The eight groups were observed by {\it XMM-Newton} for nominal
exposure times of $\sim 20$ ks. The data were processed using
{\sixrm XMMSAS} v6.0. Following the procedure outlined in Jeltema
et al. (2006), we remove primary and secondary flares and obtain
cleaned events files for the MOS1, MOS2 and PN detectors. The net
exposure times summed over the three detectors are listed in Table
\ref{table:group}. Little or no diffuse X-ray emission is detected
for the majority of the groups, indicating that these groups are
not virialized and are most likely collapsing for the first time.
Two groups MZ 9014 and MZ 4577 have some detected diffuse
emission, but at very low luminosities ($L_X\approx 10^{41}$ erg
s$^{-1}$ ; Rasmussen et al. 2006). The MZ 10451 group is X-ray
luminous (L$_X$ $\gtrsim 10^{42}$ erg s$^{-1}$), suggesting it is
more dynamically evolved than the other groups.

The {\it wavdetect} task in SAS was used to identify X-ray sources
in the fields with a detection threshold of 5 sigma. Source
detection was run on both the broad ($0.5-8$ keV) and hard ($2-8$
keV) band images. The X-ray detection limit for each group was
calculated assuming a power-law model with index $\Gamma=1.7$ and
Galactic absorption (see Table \ref{table:group}).

%We search for matches between the X-ray point sources and group
%members by a general 2$''$ match, and we also check the
%over-plotted optical and X-ray source images by eye.
The X-ray images and source lists were compared with the list of
group members\footnote{%{\bf We also match the X-ray sources with
%all the IMACS imaging sources in each field, in case that those
%potential group members without IMACS spectra have an X-ray
%counterpart. But all the few additional matches we found have
%IMACS spectra and turn out not to be group members.}
We also compare the X-ray sources with the full imaging data set.
The additional matches we found all have IMACS spectra and are
excluded as group members.} to identify possible matches. For the
hard band images, there are two matches (one in MZ 5383 and the
other in the X-ray luminous group MZ 10451). The X-ray source in
MZ 5383 is extended and has a spectrum that is well-fit
by a thermal plasma model. Hence, this object is not an X-ray AGN.
The detected galaxy in the MZ 10451 group is a point source with a very hard
X-ray spectrum and a luminosity of $L_X\sim 1.7\times 10^{41}$ erg s$^{-1}$
(in the $0.3-8$ keV band). These characteristics suggest this object
is an X-ray AGN. The host galaxy is an
elliptical and there are no obvious AGN emission-line signatures in the
optical spectrum. Therefore, the properties of this source are
similar to the X-ray AGN found in rich clusters by Martini et al.
(2006). A few additional matches to group members are found in the
broad-band images, but X-ray spectral analysis indicates
that all of these are well-described by a thermal plasma model and
thus are unlikely to be AGN. In addition, none of the five optical
AGN are detected in the X-ray images. This is consistent with the
observation that AGN selected by their emission line spectra are
often X-ray weak in contrast to AGN selected by their blue/UV
excess (Risaliti et al. 2001). Although large absorbing columns
might be responsible for the non-detection of these AGN in X-rays
(particularly for the edge-on galaxies), spectral analysis of a
sample of emission-line selected AGN suggests that most of them
are intrinsically X-ray weak (Risaliti et al. 2003).
%{\it Better provide some dust
%obscuration estimation in our five AGN.}
%We will proceed assuming that the optical AGN are intrinsically
%X-ray weak.
%As we will see in \S 4, this is a natural prediction of our
%AGN evolution picture.

\section{A Comparison of the AGN Populations of Groups and Clusters}\label{sec:comparison}

The optical/X-ray properties of the AGN in our groups are quite
different from those in rich clusters. The AGN population in
clusters is dominated by X-ray bright AGN without strong optical
emission lines (Martini et al. 2006). The majority of these AGN
occur in early-type galaxies. The one X-ray AGN in our sample also
occurs in an early-type galaxy and is a member of the only X-ray
luminous group. In many ways, such X-ray luminous groups can be
thought of as \lq\lq mini-clusters\rq\rq \ (cf. Mulchaey 2000), so the presence
of X-ray AGN in this group is perhaps not too surprising.

However, X-ray bright, optically dull AGN are
entirely absent from the less dynamically evolved groups in our sample.
It is worth noting that the overall AGN fraction in our
groups is consistent with the fraction in
cluster of galaxies ($\sim$ 7\% in our group sample vs
$\sim 5\%$ in clusters down to the same limiting magnitude of
 $M_{\rm R}=-20$; Martini et al. 2006).
Therefore, it is not the fraction of AGN that
varies significantly between these two environments, but the nature of the AGN populations:
groups are dominated by AGN with prominent optical emission lines,
while clusters are dominated by X-ray luminous, optically-dull AGN.

It is interesting to consider how the differences in the AGN
populations in poor groups and rich clusters might reflect
differences in the galaxy populations in these two environments.
As the AGN in our groups occur in late-type galaxies that are
likely encountering the group environment for the first time,
their host galaxies should retain large reservoirs of gas. In
contrast, cluster hosts are mostly early-type galaxies that are
largely void of gas and dust. Therefore, the differences in the
AGN types in groups and clusters may reflect differences in AGN
accretion rates in these two environments. Furthermore, since the
galaxy population in clusters is older than in groups, the
differences in the nature of AGN in the two environments may also
represent an evolutionary effect. In the following Section we
combine these two ideas to show how an evolving accretion rate can
explain not only our group observations, but many properties of
optically and X-ray selected AGN in general.

\section{A Scenario For AGN Accretion Evolution}\label{sec:unified_model}

The notion of an evolving accretion rate for AGN has been
considered recently by several authors (Yu, Lu \& Kauffmann 2005;
Cao 2005; Hopkins et al. 2006). Here, we propose a simple scenario
in which an evolving accretion rate leads to changes in the
accretion state of the central SMBH and therefore changes in the
observed AGN properties.

In this picture, an AGN evolves through two major stages during
its lifetime. The first stage\footnote{ The first accretion phase
may be hidden by dust obscuration until strong feedback expels
dust and clears the line of sight (e.g., Hopkins et al. 2006).} is
initiated after AGN activity is triggered by (major and/or minor)
galaxy merging. During this stage, the accretion rate is high
(probably near the Eddington rate) since there is plenty of fuel
provided by the merger; the accretion occurs via a thin $\alpha$
disk (e.g. Frank, King \& Raine 2002) and
%[i.e., $\dot{m}\equiv \dot{M}/\dot{M}_{\rm Edd}\sim 1$, where
%$\dot{M}_{\rm Edd}=2.2(M/10^8 M_\odot)M_\odot {\rm \ yr}^{-1}$ is
%the Eddington rate].
most of the luminosity is emitted in the UV/optical band with a
high radiative efficiency ($\epsilon\sim 0.1$). The ionization
flux from the accretion disk leads to prominent emission-line
regions. The SMBH is expected to assemble much of its mass during
this early phase (So\l tan 1982; Yu \& Tremaine 2002).
%; Hopkins, Narayan \& Hernquist 2006).
Observational constraints
suggest this UV/optical dominant phase typically lasts $\sim 10^6-
10^8$ yr (e.g, Martini 2004).

As gas is consumed, the accretion rate decreases, eventually
leading to a change of accretion state from thin disk accretion to
a radiatively inefficient accretion flow (RIAF, e.g., Narayan \&
Yi 1994; Abramowicz et al. 1995; Armitage 2004).% \footnote{The
%transition of a thin disk to RIAF has been successfully applied to
%the low/hard and high/soft spectral transitions in X-ray binaries
%(e.g., Esin et al. 1998). The difference here is that the
%accretion rate is more variable for X-ray binaries than for
%SMBHs.}.
The bolometric luminosity is lower during this second
phase due to both the lower accretion rate and the lower radiative
efficiency. However, as more energy is deposited in the accretion
flow rather than radiated away, the RIAF becomes hot enough to
primarily radiate in the X-ray band as inverse Compton emission
(e.g., Yuan \& Narayan 2004). Hence the AGN becomes X-ray dominant
in this later stage. Furthermore, the decrease in the UV/optical
radiation, as well as possible consumption of Broad-Line-Region
and Narrow-Line-Region clumps, leads to a substantial drop in
emission line flux. This later accretion phase can last much
longer than the initial UV/optical phase because the low accretion
rate can be easily maintained by small amounts of infalling gas
and/or relic circumnuclear medium.

This simple picture postulates that AGN are optically dominant
early on and gradually fade and change into X-ray dominant AGN. In
this scenario, optical and X-ray AGN are generically the same
population of objects, observed at different evolutionary stages.
The environmental dependence of AGN properties is then a natural
consequence of accretion evolution: AGN are preferentially
discovered as X-ray AGN in evolved systems like clusters of
galaxies, where the merger events occurred a long time ago and
rapid accretion onto the SMBH has ceased; while they are
preferentially discovered as optical AGN in dynamically young
systems like groups where the merger events have more recently
occurred\footnote{We may also expect some optical AGN in the
outskirts of clusters, where galaxies are probably infalling for
the first time.}. Some additional inferences are:
\begin{enumerate}
\item[a.] Since AGN spend a longer time in the low accretion rate, X-ray
bright phase than in the high accretion rate, optical bright
phase, the number density of X-ray AGN is larger than that of
optical AGN (Mushotzky 2004 and reference therein). Furthermore,
the number density of X-ray selected AGN should peak at later
cosmic time than optically-selected AGN (Barger et al. 2002;
Fabian 2004; Hasinger, Miyaji \& Schmidt 2005).

\item[b.] On average X-ray AGN should have lower bolometric
luminosities than optical AGN, which is consistent with the fact
that {\it Chandra} AGN sources have considerably less mean
bolometric luminosity than optically-selected AGN (e.g., Mushotzky
2004). In addition, an anti-correlation between the intrinsic
X-ray/optical luminosity ratio (or spectral hardness) and
optical/UV luminosity is expected because during the early stages
AGN are more dominant in the optical/UV band and more luminous.
This intrinsic anti-correlation is indeed observed in a recent
study of the soft X-ray properties of an optically-selected AGN
sample (Strateva et al. 2005).

\item[c.] Many authors have suggested that the X-ray emission from
optically-selected AGN is too soft to account for the hard X-ray
background (e.g., Natarajan 2004 and reference therein). This is a
natural result in our model: most optically-selected AGN are in
the optical-dominant phase and thus have relatively soft X-ray
spectra. However, sources with radiatively inefficient accretion
flows are able to produce X-ray emission with a spectral shape
similar to the hard X-ray background (Di Matteo \& Fabian 1997; Yi
\& Boughn 1998). Thus, the hard X-ray sources that dominate the
hard X-ray background are likely aged AGN which were luminous
optical AGN in the past.

\end{enumerate}

A more subtle implication is for the different clustering
properties of X-ray and optically-selected AGN. Optical AGN have a
similar clustering strength to that of galaxies at $z\lesssim
2.5$, with a comoving correlation length $\sim 6\ h^{-1}$ Mpc
(e.g., Zehavi et al. 2005; Croom et al. 2005).
%Clustering analysis of
%large samples of X-ray selected AGN has only become
%available very recently because of the {\it XMM-Newton} and {\it
%Chandra} missions (Yang et al. 2003; Basilakos et al. 2004;
%Puccetti et al. 2006).
The correlation length is larger for X-ray selected
AGN\footnote{We note, however, the current X-ray AGN clustering
samples are still very limited by the sky coverage and sample
size.} (e.g., Basilakos et al. 2004; Yang et al. 2006; Puccetti et
al. 2006), and even larger for hard X-ray selected AGN ($\sim 15\
h^{-1}$ Mpc, Puccetti et al. 2006). Naively, one might expect that
given a monotonic relationship between luminosity and mass, X-ray
AGN should reside in less massive dark matter halos since they
tend to be less luminous. Therefore, X-ray AGN are expected to
cluster less strongly than optical AGN - the opposite sense of
what is observed (e.g., Mushotzky 2004).

This puzzle is clarified in the scenario proposed above. Though
X-ray AGN are less luminous, they do {\it not} reside in less
massive dark matter halos. It is the time evolution of individual
AGN that blurs the connection between bolometric luminosity and
host dark matter halo mass. X-ray AGN were once luminous optical
AGN at higher redshift. Using a high redshift ($z\ge 2.9$) sample
from SDSS, Shen et al. (2006, submitted) have done a clustering
analysis for high redshift optical quasars and find that they are
more strongly clustered than low redshift quasars, with a comoving
correlation length comparable to that of the hard X-ray AGN.
%In our evolutionary scenario,
These once luminous optical AGN gradually fade and become less
luminous X-ray AGN at lower redshift, while their spatial
clustering properties remain.

\section{Conclusions}
We have studied the optical and X-ray properties of AGN in eight
low redshift groups of galaxies. We spectroscopically identify
$\sim 4\%$ of the group members (down to $M_{\rm R}=-18$) to be
AGN. None of these optical AGN are detected in X-rays. We detect
only one additional AGN in X-rays. This object, which shows no optical AGN signatures, is a member of the most
dynamically evolved group.
%If groups had the same fraction of
%X-ray AGN as clusters (i.e. $\sim$ 5\%), then the probability of
%detecting one or fewer X-ray AGN out of $\sim 140$ galaxies in our
%group sample would be only\footnote{We have used eqn. 15 in Gould
%et al. (2006) to estimate the probability, where we detect 35
%X-ray AGN out of $\sim 700$ galaxies in clusters (Martini et al.
%2006) while 1 X-ray AGN out of $\sim 140$ galaxies in our group
%sample.} $1\%$.
Hence unlike rich
clusters of galaxies, poor galaxy groups appear to lack a significant
population of X-ray bright, optically-dull AGN.
%We suggest this
%result can be understood if the accretion rates for SMBHs are
%higher in groups than in clusters.
%Yue: this argument is inaccurate, so I removed it.
These observations support a simple scenario where AGN
gradually evolve from optically to X-ray dominant as the
accretion rate drops. Although this simple model can
qualitatively explain many observations of X-ray and optical AGN,
a more quantitative understanding of the properties of X-ray and
optical AGN (i.e., luminosity functions, clustering properties)
will require the convolution of this time evolution for individual
AGN and many other factors (e.g., merger rate, halo mass function,
AGN lifetimes, etc.), combined with a better understanding of SMBH
accretion physics. In addition, our current group sample is still very small.
More detailed studies of both optically and X-ray selected AGN in
a range of environments will help provide further constraints on
the ideas presented here.

%In interpreting the results, we proposed a unified AGN
%evolution scenario, based on the change of accretion rate and
%accretion state of the central SMBH. This simple picture naturally
%explains many observations, and the idea of accretion rate getting
%lower and SED getting harder is quite reasonable. Within this
%framework, a more delicate modelling is feasible in the near
%future, using better constraints from optical and X-ray AGN
%luminosity functions, AGN lifetime (both in the optical and X-ray
%phase), and merger rate at different epoch, with increased
%knowledge on accretion processes (i.e., radiation efficiency,
%accretion rate, emission spectral energy distribution).

\acknowledgments We are grateful to Greg Walth for help with the
data reduction. We also thank Luis Ho, Tesla Jeltema, Dan Kelson,
Juna Kollmeier, Paul Martini and Michael Strauss for useful
discussions. YS thanks Carnegie Observatories for hospitality
during the course of this work. JSM acknowledges support from
NASA grant NNG04GC846.

%\clearpage

%\begin{figure}
%\centering
%\includegraphics[scale=0.5]{f1.eps}
%\caption{The BPT contour diagram for line ratios using a sample of 20,000 galaxies from the SDSS.
%    The left panel shows the traditional BPT diagram using flux ratios, while the right panel shows
%    the alternative diagram using EW ratios. The red line is taken from eqn. 5 of
%    Kewley et al. 2001. Points to the right of the red line are identified as AGN. The
%    two data points with error bars (the $x-$axis errors are too small to see)
%    are identified optical AGN in MZ 4577 with all the four lines measurable.}
%\label{BPT}
%\end{figure}

%\clearpage

\end{document}